\documentstyle[epsfig,12pt]{article}  
\newcommand{\beq}{\begin{equation}}
\newcommand{\eeq}{\end{equation}}
\newcommand{\bea}{\vspace{0.25cm}\begin{eqnarray}}
\newcommand{\eea}{\end{eqnarray}}

\newcommand{\r}{\mbox{{\boldmath
$\rho$}}}

\newcommand{\ro}{\mbox{{\boldmath
$\rho$}}}

\newcommand{\rr}{\mbox{{\boldmath
$\rho$}}}

\newcommand{\qb}{\mbox{{\bf
q}}}
\newcommand{\fb}{\mbox{{\bf
f}}}
\newcommand{\pb}{{{\bf p}}}

\newcommand{\Hb}{\mbox{{\bf
H}}}
\newcommand{\Eb}{\mbox{{\bf
E}}}
\newcommand{\Fb}{\mbox{{\bf
F}}}
\newcommand{\Gb}{\mbox{{\bf
G}}}

\def\lsim{\mathrel{\rlap{\lower4pt\hbox{\hskip1pt$\sim$}}
    \raise1pt\hbox{$<$}}}         
\def\gsim{\mathrel{\rlap{\lower4pt\hbox{\hskip1pt$\sim$}}
    \raise1pt\hbox{$>$}}}         
\begin{document}
\vspace*{-2cm}
 
\bigskip

\begin{center}

\renewcommand{\thefootnote}{\fnsymbol{footnote}}

  {\Large\bf
Parton energy loss in glasma
}
\\
\vspace{.7cm}
\renewcommand{\thefootnote}{\arabic{footnote}}
\medskip
  {\large
  P.~Aurenche$^a$ and B.G.~Zakharov$^{b}$}
  \bigskip

{\it
$^{a}$
LAPTH, Universit\'e de Savoie, CNRS,\\
BP 110, F-74941, Annecy-le-Vieux Cedex, France\\
$^{b}$L.D. Landau Institute for Theoretical Physics,
        GSP-1, 117940,\\ Kosygina Str. 2, 117334 Moscow, Russia\\
\vspace{1.7cm}}
  {\bf
  Abstract}
\end{center}
{
\baselineskip=9pt
We study the synchrotron-like gluon emission in $AA$-collisions
from fast partons due to interaction with the coherent glasma color fields.
Our results show that for RHIC and LHC conditions the contribution
of this mechanism to parton energy loss 
is much smaller than 
the radiative energy loss in the plasma phase. 
\vspace{.5cm}
}

\section{Introduction}
It is widely believed that the observed 
large suppression of high-$p_{T}$ hadrons
in $AA$-collisions at RHIC and LHC (the so-called jet quenching (JQ))
is due to parton energy loss in the quark-gluon plasma (QGP).
The observation of this phenomenon and 
successful 
hydrodynamic description of hadron 
production at low $p_{T}$ are the strongest arguments for formation 
of the QGP in  heavy-ion collisions at RHIC and LHC.
In pQCD partons lose energy mostly through the 
induced gluon emission due to multiple scattering
in the QGP \cite{BDMPS,LCPI,BSZ,W1,GLV1,AMY}, and a small
fraction of the energy loss comes from the elastic collisions 
 \cite{Ecol1,Ecol2}. 
At RHIC and LHC, the entropy of the QGP required to describe the 
observed suppression of high-$p_T$ hadrons, within the pQCD calculations 
of the radiative
and collisional energy losses 
\cite{RAA04,RAA08,RAA11}, turns out to be 
qualitatively consistent with that
obtained within the hydrodynamic simulations from the low-$p_T$ data.
This consistency  looks very encouraging. However, one must keep in mind that 
the theoretical uncertainties both in 
the pQCD calculations of the JQ
and in the hydrodynamic simulations  are rather large.
On the JQ side there remain open questions on additional 
mechanisms of the parton energy loss which can modify the picture 
with the dominating radiative energy loss in the QGP phase.

One of the potentially important energy loss mechanisms
is the synchrotron-like 
gluon emission 
in the strong classical color fields in the initial pre-equilibrium stage
of $AA$-collisions \cite{KMW} predicted within 
the Color Glass Condensate (CGC) model 
\cite{MV1,MV2,Leonidov_CGC_rev,Gelis_CGC_rev}.
This stage (termed glasma \cite{term_gl})
is now under active investigation
(see \cite{McLerran_gl_rev,Lappi_gl_rev} and references therein). However,
its role in parton energy loss still has not been studied. 
The evaluation of the classical Yang-Mills fields in $AA$-collisions within
the CGC model shows that just after
the collision of the Lorentz contracted nuclei 
a system of the color flux tubes with the longitudinal boost invariant 
color electric and color
magnetic fields (with $|E_z|\approx|B_z|$) should be produced \cite{Lappi}. 
The typical transverse coherence length for the color fields in this phase
is $\sim 1/Q_{s}$, where
$Q_s$ ($\sim 1-1.5$ GeV for RHIC and LHC conditions \cite{Lappi_qs}) is 
the saturation scale of the nuclear parton distributions.
At the proper time $\tau\sim 1/Q_s$ the strength of the glasma electric 
and magnetic longitudinal fields is $gE_{z}\sim gB_{z}\sim Q_{s}^{2}$
\cite{Lappi}. 
The transverse fields, which are absent at $\tau=0$, 
are generated at later times and become close to the longitudinal 
ones at $\tau\gsim 1/Q_s$ \cite{Lappi,Itakura}. At such times
the glasma energy density decreases $\propto 1/\tau$.
At later times the thermalization of the glasma color fields 
should lead to formation of the equilibrium
QGP, but the detailed mechanism for the thermalization remains
unclear. Qualitative analyses say that the glasma thermalization goes probably 
via instabilities of the color flux tubes 
rising quickly at $\tau\gsim 1/Q_s$ \cite{RV,Itakura,Iwazaki}. 
They should lead to a fast randomization of the boost invariant 
glasma color fields at the time scale about 2--3 units of $1/Q_{s}$. 
This thermalization time agrees qualitatively with the hydrodynamic 
fits of the data 
on $AA$-collisions which favor the thermalization 
at $\tau_{0}\sim 0.4-1$ fm \cite{Heinz_tau1,Heinz_tau2}.

For RHIC-LHC conditions the typical Lorentz force acting on a fast 
parton crossing
the glasma turns out to be very large $\sim Q_{s}^{2}\sim 5-10$ GeV/fm.
It is about 10--20 times that for the Debye screened color center in the QGP 
$\sim \alpha_{s} m_{D}^{2}$ (if one takes $\alpha_{s}\sim 0.3$ and
$m_{D}\sim 0.5$ GeV).
It raises a natural question about the impact of the glasma
on the radiative energy loss. 
If one ignores the interaction of fast partons with 
the pre-equilibrium phase at all, then for jets with 
energies $\lsim 100$ GeV the parton showering at 
$\tau\lsim \tau_{0}\sim 0.5-1$ fm can be viewed as almost pure 
DGLAP shower, and the interference of the DGLAP stage 
with the induced gluon emission in the bulk of the QGP at $\tau\gsim \tau_0$
should be relatively small \cite{RAA08}. Thus in this energy range
the question of the glasma role in the JQ is equivalent to the 
question to what extent the glasma modifies the DGLAP gluon emission.  
The purpose of the present
paper is to give a qualitative estimate of this effect. 

\section{Formulation of the model and basic formulas for the synchrotron-like 
gluon emission}

An accurate calculation of the synchrotron-like gluon emission due to 
interaction of fast partons with the glasma is presently impossible.
It would require a detailed understanding of the decay and thermalization
of the 
glasma color flux tubes which remain unclear.
However, since the strength
of the color fields in the flux tubes decreases quickly,
the transverse momentum which a fast parton 
gets in the glasma should mostly come from interaction with 
the coherent color field of the first crossed color flux tube.
The effect of the random transverse momentum kicks which a fast parton
undergoes in the glasma at later times 
should be small due to weakness of the fields and the 
destructive interference of the chaotic contributions from 
different color tubes.
For the above reasons as a first step in understanding the 
influence of the glasma on 
parton energy loss it seems reasonable to consider a simple model with
a uniform time-dependent color field which acts only for a limited 
range of $\tau$ about 2--3 units of
$1/Q_s$. 

So we will consider the synchrotron-like gluon emission from 
a fast parton (we choose the $z$-axis along the initial parton momentum) in a
slab of thickness $L\sim (2-3)/Q_{s}$ with the 
transverse chromoelectric, $\Eb_a$, and
chromomagnetic, $\Hb_{a}$, fields (hereafter $a$ denote the color index).
We will ignore the longitudinal (along the jet) components of the
color fields since similarly to the photon emission in QED 
their effect should be small for relativistic partons. 
For simplicity we assume that both the electric and magnetic fields are
oriented to the same direction in the color space. 
It is enough to consider the color fields with nonzero 
components only in the Cartan subalgebra, i.e., for $a=3$ and $a=8$ for 
the $SU(3)$ color group. 

We present our formulas for a fast quark produced in a hard process 
at central rapidity $y=0$ (so the initial quark momentum and the $z$-axis
of our coordinate system are perpendicular to the $AA$-collision axis).
We take $z=0$ for the production point. The starting point of our 
analysis is similar
to the case of the synchrotron-like energy loss spectrum  for a fast parton
propagating in an infinite uniform color field addressed in \cite{synQCD}.
We write the $S$-matrix element of the $q\to gq'$ transition
as (we omit the color factors and indices)
\beq
\langle gq'|\hat{S}|q\rangle=-ig\int\! d^{4}y 
\bar{\psi}_{q'}(y)\gamma^{\mu}A_{\mu}^{*}(y)\psi_{q}(y)\,,
\label{eq:10}
\eeq
where $\psi_{q,q'}$ are the wave 
functions of the initial quark and final quark, $A_{\mu}$ is the wave function
of the emitted gluon. 

For our choice of the external color field (which has only 
$a=3$ and $a=8$ nonzero components) it does not
change the quark color state. However, it is not the case for gluons, which
are rotated in the color space as they move in the color field.
The interaction of the gluons with the background
color field can be diagonalized by introducing the gluon fields having definite
color isospin, $Q_{A}$, and color hypercharge, $Q_{B}$, (we will describe 
the color charge by the two-dimensional vector $Q=(Q_A,Q_B)$).
In terms of the usual gluon vector potential, $A_{a}$, the diagonal color gluon
states read
(the Lorentz indices are omitted)
$X=(A_{1}+iA_{2})/\sqrt{2}$ ($Q=(-1,0)$),
$Y=(A_{4}+iA_{5})/\sqrt{2}$ ($Q=(-1/2,-\sqrt{3}/2)$),
$Z=(A_{6}+iA_{7})/\sqrt{2}$ ($Q=(1/2,-\sqrt{3}/2)$).
The neutral gluons $A=A_{3}$ and $B=A_{3}$ with $Q=(0,0)$,
to leading order in the coupling
constant, do not interact with the background field at all, and the emission of
these gluons are similar to the photon radiation in QED.

We evaluate the matrix element (\ref{eq:10}) within the small angle
approximation, i.e., we assume that for each parton its transverse momentum is
small compared to the longitudinal one. 
We write each quark wave function in the form
\beq
\psi_{i}(y)=\frac{\exp[-iE_{i}(t-z)]}{\sqrt{2E_{i}}}
\hat{u}_{\lambda}
\phi_{i}(z,\r)\,,
\label{eq:20}
\eeq
where the four-vector $y$ is defined as
$y^{\mu}=(t,\r,z)$ (hereafter the bold vectors denote the
transverse vectors),
$\lambda$ is the quark helicity,
$\hat{u}_{\lambda}$ is the Dirac spinor operator. 
In the small angle approximation the $z$-dependence of the transverse wave 
functions
$\phi_{i}$ is governed by the two-dimensional 
Schr\"odinger equation 
\beq
i\frac{\partial\phi_{i}(z,\r)}{\partial
z}={\Big\{}\frac{
(\pb-gQ_{n}\Gb_{n})^{2}
+m^{2}_{q}} {2E_{i}}
+U_{i}(z,\r){\Big\}}
\phi_{i}(z,\r)\,
\label{eq:30}
\eeq
with the potential
\beq
U_{i}(z,\r)=gQ_{n}^{i}[G^{0}_{n}(z,\r)-G^{3}_{n}(z,\r)]\,.
\label{eq:40}
\eeq
Here $G^{\mu}_{n}$ denotes the external vector potential 
(the superscripts are the Lorentz indexes and 
$n=1,2$ correspond to the $A$ and $B$ 
color components in the Cartan subalgebra), $Q_{n}^{i}$ is the 
quark color charge.

We take the external vector potential in the form 
$G^{0}_{n}=-\r\cdot \Eb_{n}$, $\Gb_{n}=0$, and
$G^{3}_{n}=[\Hb_{n}\times \r]^{3}$, where
$\Eb_{n}$ and $\Hb_{n}$ are the electric and magnetic fields. 
In this case the kinetic term in (\ref{eq:30}) does not contain the
external vector potential at all, and  
the potential $U_{i}$ can be written as
\beq
U_{i}(z,\r)=-\Fb_{i}\cdot\r\,,
\label{eq:50}
\eeq
where $\Fb_{i}$ is the Lorentz force.
The wave function of the emitted
gluon can be represented in a similar way.
In our numerical computations
for the quark mass we take $m_{q}=0.3$ GeV.
However, the value of the quark mass is not very important. 
For the gluon mass we take $m_{g}=0.75$ GeV. This value was obtained  
from the analysis of the low-$x$ proton structure function $F_{2}$ within 
the dipole BFKL equation
\cite{NZ_HERA}. It agrees well with the natural infrared cutoff 
for gluon emission $m_{g}\sim 1/R_{c}$, where 
$R_{c}\approx 0.27$ fm is the gluon correlation radius in the QCD vacuum
\cite{shuryak1}. 

The solution of (\ref{eq:30}) can be taken in the form
\beq
\phi_{i}(z,\r)=\exp{\left\{i\pb_{i}(z)\r-
\frac{i}{2E_{i}}\int_{0}^{z}dz'[\pb^{2}_{i}(z')+m^{2}_{q}]\right\}}\,.
\label{eq:60}
\eeq
Here the transverse momentum $\pb_{i}(z)$ is the solution to 
the classical parton equation of motion in the impact parameter plane
\beq
\frac{d\pb_{i}}{dz}=\Fb_{i}(z)\,.
\label{eq:70}
\eeq

By using (\ref{eq:10}), (\ref{eq:20}), (\ref{eq:60}) one can obtain 
\bea
\langle gq'|\hat{S}|q\rangle=-ig
(2\pi)^{3}\delta(\omega+E_{q'}-E_{q})
\int_{0}^{\infty}dz V(\qb(z),\{\lambda\})
\delta(\pb_{g}(z)+\pb_{q'}(z)-\pb_{q}(z))\nonumber\\
\times
\exp{\left\{i\int_{0}^{z} dz'
\left[
\frac{\pb^{2}_{g}(z')+m^{2}_{g}}{2\omega}
+\frac{\pb^{2}_{q'}(z')+m^{2}_{q}}{2E_{q'}}
-\frac{\pb^{2}_{q}(z')+m^{2}_{q}}{2E_{q}}
\right]\right\}}\,,
\label{eq:80}
\eea
where $\omega$ is the gluon energy, $V$ is the vertex factor, 
$\{\lambda\}$ is the set of the parton
helicities. For the transition conserving
quark helicity 
$V(\qb,{\lambda})=
-iC[2\lambda_{q}x+(2-x)\lambda_{g}]
[q_{x}(z)-i\lambda_{g}q_{y}(z)]
/[x\sqrt{2(1-x)}]$, and for the spin-flip
case $V=ixm_{q}C(2\lambda_{q}\lambda_{g}+1)/\sqrt{2(1-x)}$. Here
$\qb(z)=\pb_{g}(z)(1-x)-\pb_{q'}(z)x$, $x=\omega/E_{q}$,
and $C=\lambda_{fi}^{a}\chi_{a}^{*}/2$ is the color factor ($i,f$ are the color 
indexes of the initial and final quarks, $\chi_{a}$ is the color wave 
function of the emitted gluon).
Due to 
the color charge conservation $\Fb_q=\Fb_g+\Fb_{q'}$. For this reason,
the argument of the momentum $\delta$-function on the right-hand part 
of (\ref{eq:80}) does not depend on $z$,
and can be replaced by $\pb_{g}^{+}+\pb_{q'}^{+}-\pb_{q}^{+}$,
where  
$\pb_{i}^{+}=\pb_{i}(\infty)$.
Then, noting that the expression in the square brackets in (\ref{eq:80})
can be rewritten as $[\qb^{2}(z')+\epsilon^{2}]/2M$ with
$\epsilon^{2}=m_{q}^{2}x^{2}+(1-x)m_{g}^{2}$ and $M=E_{q}x(1-x)$, 
the matrix element (\ref{eq:80}) can be rewritten as
\beq
\langle gq'|\hat{S}|q\rangle=-i
(2\pi)^{3}\delta(\omega+E_{q'}-E_{q})
\delta(\pb_{g}^{+}+\pb_{q'}^{+}-\pb_{q}^{+})T\,,
\label{eq:90}
\eeq
\beq
T=
g\int_{0}^{\infty}dz V(\qb(z),\{\lambda\})
\exp{\left\{i\int_{0}^{z} dz'
\left[\frac{\qb^{2}(z')+\epsilon^{2}}{2M}
\right]\right\}}\,.
\label{eq:100}
\eeq

With the help of the standard Fermi golden rule the gluon distribution
in terms of the amplitude $T$ can be written as
\beq
\frac{dN}{d\omega d\qb}=\frac{|T|^{2}}{8(2\pi)^{3}E_{q}^{3}x(1-x)}\,.
\label{eq:110}
\eeq
Hereafter we do not show the averaging over the initial
and summing over the final color/spin states. 
In the absence of the external color field the momentum
$\qb(z)$ in (\ref{eq:100}) does not vary with $z$ and equals to  
the final momentum at $z=\infty$ (we will denote it $\qb)$. 
In this case (\ref{eq:100}) gives the vacuum amplitude
\beq
T_{v}=gV(\qb,\{\lambda\})
\frac{2iM}{\qb^{2}+\epsilon^{2}}\,.
\label{eq:120}
\eeq
Substituting (\ref{eq:120}) into (\ref{eq:110}) we obtain 
the usual vacuum LO pQCD distribution for the $q\to g q$ splitting
(hereafter we neglect the spin-flip part of the vertex ($\propto m_{q}$)
which gives a negligible contribution)
\beq
\frac{dN_{v}}{d\omega d \qb}=
\frac{2 C_{F}\alpha_{s}}{\pi^{2}xE_{q}}
\Big(1-x+\frac{x^{2}}{2}\Big)\frac{\qb^{2}}{(\qb^{2}+\epsilon^{2})^{2}}\,.
\label{eq:130}
\eeq

For a nonzero external field we write the amplitude $T$ as a sum
\beq
T=T_{v}+T_{s}\,,
\label{eq:140}
\eeq
where $T_{s}$ describes the correction
due to the external field.
Since the momentum $\qb$ varies with $z$ only at $z<L$,
from (\ref{eq:100}) one can obtain 
\bea
T_{s}=
g\int_{0}^{L}dz V(\qb(z),\{\lambda\})
\exp{\left\{i\int_{0}^{z} dz'
\left[\frac{\qb^{2}(z')+\epsilon^{2}}{2M}
\right]\right\}} - (\qb(z)\to \qb)\,.
\label{eq:150}
\eea

In terms of $T_{v,s}$ the synchrotron-like correction to the LO gluon 
spectrum (\ref{eq:130}) reads
\beq
\frac{dN_{s}}{d\omega d\qb}=\frac{2\mbox{Re} (T_{v}T^{*}_{s})
+|T_{s}|^{2}}{8(2\pi)^{3}E_{q}^{3}x(1-x)}\,.
\label{eq:160}
\eeq
The glasma correction to the $\omega$-spectrum can be obtained 
from (\ref{eq:160})
by integrating over the transverse momentum $\qb$. Note that 
contrary to the LO vacuum spectrum (\ref{eq:130}),
which $\propto 1/\qb^{2}$ at large $\qb^{2}$ and 
without a constraint on $\qb^{2}$ gives a logarithmically 
divergent $\omega$-spectrum,
the $\qb$-integral for 
the correction term is convergent at large $\qb^{2}$. 

The $\omega$-spectrum can also be obtained within
the light-cone path integral (LCPI) approach \cite{LCPI} formulated
in the impact parameter space. The LCPI formalism has originally been developed
for gluon emission induced by multiple parton scattering. But 
it applies to the synchrotron-like
gluon emission as well. In this case the LCPI gives for the $\omega$-spectrum
\beq
\frac{dN_{s}}{d\omega}\! =2{\rm Re} \int^{\infty}_{0}\! dz_1\! 
\int^{\infty}_{z_1}\!
dz_2 \hat{g} \left[{\cal{K}} (\rr_2 , z_2 | \rr_1 , z_1)\right.
 -\left. {\cal{K}}_{0} 
(\rr_2 , z_2 | \rr_1 , z_1 )\right]{\Big|}_{\rr_1 = \rr_2 = 0}\,.
\label{eq:170}
\eeq
Here 
$\hat{g}$ is the 
vertex operator given by
\bea
\hat{g}=\frac{\alpha_{s}}{8E_{q}^{3}x(1-x)}\sum_{\{\lambda\}}
V(-i{\partial}/{\partial \rr_{1}},\{\lambda\})
V^{*}(-i{\partial}/{\partial \rr_{2}},\{\lambda\})\nonumber\\
=\frac{|C|^{2}\alpha_{s}}{E_{q}^{3}x^{3}(1-x)^{2}}\left(1-x+\frac{x^{2}}{2}\right)
\frac{\partial}{\partial \rr_{1}}\cdot
\frac{\partial}{\partial \rr_{2}}\,,
\label{eq:180}
\eea
$\cal{K}$ is the Green's function of the Schr\"odinger equation
with the Hamiltonian
\beq
\hat{H}=
-\frac{1}{2M}\,
\left(\frac{\partial}{\partial \rr}\right)^{2}
-\fb\cdot\ro +\frac{\epsilon^{2}}{2M}\,,
\label{eq:190}
\eeq
and ${\cal{K}}_{0}$ is the Green's function for the Hamiltonian (\ref{eq:190}) 
with $\fb=0$. 
The Green's function for the Hamiltonian (\ref{eq:190}) is known explicitly 
(see, for example, \cite{FH})
\beq
{\cal{K}} (\rr_2 , z_2 | \rr_1 , z_1)=\frac{M}{2\pi i\Delta z}
\exp{[i S_{cl}]}\,,
\label{eq:200}
\eeq
where $\Delta z=z_2-z_1$ and $S_{cl}$ is the classical action for the
Hamiltonian
(\ref{eq:190}) given by
\bea
S_{cl}=-\frac{\Delta z\epsilon^{2}}{2M}+\frac{M}{2\Delta z}
\left[(\rr_2-\rr_1)^{2}+\frac{2}{M}\int_{z_1}^{z_2} dt\,
\rr_2\cdot  \fb(t)(t-z_1)\right. \nonumber\\
+\left.\frac{2}{M}\int_{z_1}^{z_2} dt\, \rr_1\cdot \fb(t)(z_2-t)
-\frac{2}{M^2}\int_{z_1}^{z_2} dt \int_{z_1}^{t} ds\,
\fb(t)\cdot \fb(s)(z_2-t)(s-z_1)\right]\,.
\label{eq:210}
\eea
One can show that the LCPI formula (\ref{eq:170}) being rewritten in the 
momentum 
space reproduces the $\omega$-spectrum corresponding to (\ref{eq:160}).
Note that the subtraction of the ${\cal{K}}_{0}$ term 
in (\ref{eq:170}) corresponds to subtraction of the vacuum
part from (\ref{eq:110}) for getting the synchrotron correction
(\ref{eq:160}).

Note however that for RHIC and LHC conditions in our approach
the energy loss spectrum can not be calculated accurately
at gluon energy $\omega\lsim 5$ GeV.
Our numerical calculations show that for such soft gluons the large 
angle emission
becomes important and the integration over the transverse momentum
can not be performed accurately using the formulas based on the small angle
approximation. For this reason our results on the energy loss
can only be treated as qualitative estimates, rather than 
quantitative predictions.

We presented the formulas for the $q\to gq'$ transition. For the 
purely gluonic $g\to gg$ transition the calculations are similar. 

\section{Numerical results}

For numerical calculations we should fix the $z$-dependence of the 
Lorentz force. 
As we have said, we assume that the external field acts a finite
time/length $\sim 1/Q_{s}$.
In our numerical calculations we take $L=2/Q_{s}$.  
We take $Q_{s}=1$
for RHIC and $Q_{s}=1.4$ for LHC. This gives 
$L(\mbox{RHIC})\approx 0.4$ fm and 
$L(\mbox{LHC})\approx 0.28$ fm.
To fix the $z$-dependence of the Lorentz force, which  a fast
parton feels  at $\tau\!=\!z<L$, we use the $\tau$-dependence of the glasma energy density
obtained in the lattice simulations by Lappi \cite{Lappi}.
For our calculations we need only the field components transverse
to the initial parton momentum. 
At $\tau\ll 1/Q_{s}$, when the electric and magnetic  
fields are almost parallel to the $AA$-collision axis 
\cite{McLerran_gl_rev,Lappi_gl_rev}, 
the Lorentz force acting on a fast parton is purely transverse to the 
parton momentum. At such times for a unit color charge 
$F^{2}=2g^{2}\varepsilon$, where $\varepsilon=(E^{2}+H^{2})/2$ is 
the color field energy density.
However, this relation is invalid at 
$\tau\gsim 1/Q_{s}$
when the contribution of the transverse (to the $AA$-collision axis) 
components of the color fields to the energy density becomes approximately 
equal to that from the components along the beam axis \cite{Lappi}.
Since only half of these transverse fields squared 
contribute to the Lorentz force squared which we need (transverse to
the jet direction) we can write in this regime
$F^{2}=g^{2}3\varepsilon/2$. We will use this relation to determine the 
Lorentz force  from the energy density 
obtained in \cite{Lappi}. 

The results of \cite{Lappi} were presented in terms of the coupling
constant $g$ and the mass parameter $\mu$ of the CGC model \cite{MV1,MV2} 
related to the saturation scale by the relation $Q_{s}\approx 6 g^{2}\mu/4\pi$.
Extrapolating to the continuum limit Lappi obtained 
$\varepsilon(\tau=1/g^{2}\mu)=0.26(g^{2}\mu)^{4}/g^{2}$, and  at $\tau\gsim
1/g^{2}\mu$ the energy density behaves as $\sim 1/\tau$. We use
the above value of $\varepsilon(\tau=1/g^{2}\mu)$ to fix the normalization, 
and the $\tau$-dependence of $\varepsilon$ was obtained using
the results for $g^{2}\varepsilon/(g^{2}\mu)^{4}$ 
presented 
in Fig.~3 of \cite{Lappi}. 
In \cite{Lappi} the glasma energy density has been calculated
taking $g=2$,  and 
$g^{2}\mu=2$ GeV for RHIC and $g^{2}\mu=3$ GeV for LHC.
Note that it gives the values of the 
glasma energy density at $\tau=1/g^{2}\mu$ which match 
very well to the QGP energy density extracted within the Bjorken model with 
the longitudinal 
expansion \cite{Bjorken} using the entropy/multiplicity ratio
$dS/dy\Big{/}dN_{ch}/d\eta=7.67$ 
obtained in \cite{BM-entropy} and experimental multiplicities for 
the central $Au+Au$ collisions at $\sqrt{s}=200$ GeV  at RHIC
and $Pb+Pb$ collisions at $\sqrt{s}=2.76$ TeV at LHC. In terms of the QGP
temperature it corresponds to  
$T_{0}\approx 300$ MeV for RHIC, and $T_{0}\approx 400$ MeV for LHC 
at $\tau_{0}=0.5$ fm. We use these parameters to fix 
the Lorentz force. For the coupling constant in the amplitude of the
$q\to qg$ transition (\ref{eq:100}) we also take $g=2$, i.e. 
$\alpha_s\approx 0.318$.
Our numerical evaluations show that the results
for the external fields in the $A$ and $B$ color states are very similar.
Below we present the gluon emission spectra defined as a half-sum of  
the $A$ and $B$ contributions.

To illustrate the transverse momentum dependence of the gluon
emission in Fig.~1 we plotted the distribution 
$dN_{s}/d\omega d\qb$ scaled by its value at $\qb=0$
for $\qb$ along the $x$ and $y$ axes 
(we take the $x$-axis parallel to the Lorentz force,
and the $y$-axis perpendicular to the Lorentz force) 
for a quark with energy $E=50$ GeV and $\omega=2,$ 5, 10, and 25 GeV
for RHIC and LHC conditions. The $q_{x}$-dependence in Fig.~1 is shown
for the distribution averaged over two directions of the Lorentz force. 
Fig.~1 shows that the glasma correction has a complicated $\qb$-dependence
with a region where it is negative. 
The appearance of the negative $dN_{s}/d\omega d\qb$
shows that a significant contribution comes from the interference of 
$T_{v}$ and $T_{s}$ amplitudes in (\ref{eq:160}). 
This says that the finite-size effects are very important for RHIC and LHC 
conditions (see below). 
From Fig.~1
one sees that the $\qb$-distribution
is rather broad, and large angle gluon
emission is clearly important at $\omega\lsim 5$ GeV. 
However, our approach, based on the small angle approximation, 
becomes inapplicable at the angles $\gsim 1$. For this reason 
for the $p_{T}$-integrated energy loss spectrum $dN_{s}/d\omega$ 
our approach can give only a rough estimate of the magnitude of the 
glasma contribution at $\omega \lsim 5$ GeV.
The observed significant contribution of the large angle gluon emission 
is somewhat unexpected. Indeed,  
one might naively expect that the typical transverse momenta 
should be $\sim Q_{s}$ (since the Lorentz force is $\sim Q_{s}^{2}$
and the path length is $\sim 1/Q_{s}$).

We evaluated the $dN_{s}/d\omega$ for two different prescriptions: (a) without any
constraint on the transverse momentum (ignoring the fact that our formulas are
inapplicable at large angles), and (b) imposing the kinematical constraint 
$|\qb|<\mbox{min}(\omega,E-\omega)$ (in this region the emission angles $\lsim
1$ and our formulas are qualitatively reasonable).
In Fig.~2 we plotted $\omega dN_{s}/d\omega$ for RHIC (left) and LHC 
(right) conditions for (a) (upper panels) and (b) (lower panels) prescriptions.
One can see that the energy loss is concentrated at
$\omega \lsim 5$ GeV. In this region the kinematical constraint
suppresses  the distribution by a factor $\sim 5$.
It seems reasonable to assume that the results for the prescriptions (a) 
and (b) 
define an approximate uncertainty band of the energy loss due to the glasma
color fields. 

The results presented in Fig.~2 show that the glasma should not affect
strongly the radiative energy
loss for RHIC and LHC conditions. Indeed, even without the kinematical
constraint the distributions shown in Fig.~2 turn out to be smaller by a
factor $\sim 10-20$ than the contribution of the induced gluon emission
in the QGP calculated in \cite{RAA04}.
It is also seen from calculation of the total energy loss 
$\Delta E=\int_{\omega_{min}}^{\omega_{max}} d\omega \omega dN_{s}/d\omega$.
Taking $\omega_{min}=m_{g}$ and
$\omega_{max}=E_{q}/2$ without kinematical constraint 
we obtained for RHIC (LHC) $\Delta E\approx$184 (320), 234 (414), and
276 (495) MeV for $E_{q}=10$, 20, and 50
GeV, respectively. 
For the same set of $E_{q}$ imposing the kinematical constraint 
gives the values $\Delta E\approx$48 (63), 49 (59), and
45 (39) MeV. These values were obtained for $m_{g}=0.75$ GeV. To estimates
the infrared sensitivity we also performed calculations 
for  $m_{g}=0.4$ GeV. In this case we obtained a rather small $\sim 10-20$\%
increase of $\Delta E$. We also studied the sensitivity to the thickness
$L$. If we choose $L=3/Q_{s}$ (which is probably too large,
since the glasma fields should become chaotic at such times)
the energy loss increases by at most a 
factor $\sim 1.5-2$.
The  obtained  values of $\Delta E$ are clearly small compared to that from the
induced gluon emission in the QGP which gives 
$\Delta E\sim 5-15\,$$(10-30)$ GeV 
for RHIC (LHC) conditions at $E_{q}\sim 10-50$ GeV \cite{Ecol1}.

Thus our calculations demonstrate that, despite a huge Lorentz force 
acting on fast partons in the glasma, the effect of the glasma color
tubes on parton energy loss turns out to be rather small.
This smallness is due partly to strong finite-size effects.
The point is that, similarly to the induced gluon emission due to
multiple scattering,
the synchrotron-like gluon emission is strongly suppressed when
the parton path length in the field is of the order of (or smaller)
the gluon formation length, $L_{f}$ 
\footnote{The formation length of the synchrotron gluon emission 
for a constant $\fb$ 
can be estimated using the formula 
$L_{f}\sim \mbox{min}(L_{1},L_{2})$, where $L_{1}=2\mu/\epsilon^{2}$ and
$L_{2}=(24\mu/\fb^{\,2})^{1/3}$ \cite{synQCD}.}. 
Physically this suppression is connected
with the fact that the initial hard parton at $\tau=0$ does not have a 
formed cloud of virtual gluons (with virtuality $Q\lsim E$), and
its splitting to a two parton state is simply impossible before formation
of the higher Fock states in its wave function.
To see more clearly how strong the finite-size effects are 
we calculated the finite-size suppression factor, $S$.
We define it
as a ratio 
$S=dN_{s}/d\omega\Big/dN_{s}^{'}/d\omega$, where $dN_{s}/d\omega$ is 
our $\omega$-spectrum without kinematic constraint (\ref{eq:170}),
and $dN_{s}{'}/d\omega$ is the spectrum 
in the approximation of zero formation length given by
$dN_{s}^{'}/d\omega=\int_{0}^{L} dz dN_{s}(z)/dz d\omega$.
Here $dN_{s}(z)/dzd\omega$ is the  
probability distribution of the synchrotron-like gluon emission per unit
length (also without kinematic constraint) for a uniform field 
equals to the real local field at position $z$.
In the LCPI formulation the corresponding expression reads
\beq
\frac{dN_{s}(z)}{dzd\omega}\! =2{\rm Re} 
\int^{\infty}_{z_{1}}\!
dz_{2} \hat{g} \left[{\cal{K}} (\rr_2 , z_{2} | \rr_1 , z_{1})\right.
 -\left. {\cal{K}}_{0} 
(\rr_2 , z_{2} | \rr_1 , z_{1} )\right]{\Big|}_{\rr_1 = \rr_2 = 0,\,z_{1}=z}\,,
\label{eq:220}
\eeq
which should be calculated with the 
Hamiltonian (\ref{eq:190}) but for a constant $\fb$ equals to the 
real Lorentz force at position $z$. The spectrum  
(\ref{eq:220}) can be written in terms of the Airy function \cite{synQCD}.
For a very large slab with a smooth $z$-dependence of the Lorentz 
force, when at each $z$ the local gluon formation length satisfies the
inequality $L_{f}\ll z$, the local radiation rate 
$dN_{s}(z)/dzd\omega$ becomes well defined. In this regime
the factor $S$ should become close to unity.
Thus, the deviation of $S$ from unity characterizes the strength of 
the finite-size
effects.
In Fig. 3 we plotted the suppression factor $S$
for RHIC and LHC conditions for several quark energies.
From Fig. 3 one can see that in the region $\omega\lsim 5$ GeV dominating energy loss
the finite-size
effects suppress the radiation rate by a factor $3-5$.
Such strong suppression is not surprising since in this region 
the gluon formation length is of the order of the glasma life time.

We presented the numerical results for gluon emission from a fast quark.
For a fast gluon the gluon emission spectrum $dN_{s}/dxd\qb$ is 
symmetric in $x\leftrightarrow 1-x$. At $x\lsim 0.5$ it is very similar 
in form to that for a quark, but enhanced by a factor $\sim C_{A}/C_{F}=9/4$.

\section{Summary}

In this paper we studied the synchrotron-like gluon emission induced by 
interaction of fast partons with the coherent color fields
in the glasma phase which should exist in $AA$-collisions at $\tau\sim
1/Q_{s}$, before formation of the QGP. 
We modeled the glasma by a time-dependent Lorentz force
acting on fast partons at $\tau\lsim (2-3)/Q_{s}$. 
We evaluated the synchrotron-like gluon emission
using a quasiclassical formalism
based on the small 
angle approximation. But our numerical calculations show 
a significant contribution of the large
angle region for gluons with $\omega\lsim 5$ GeV 
which are important for the total energy loss.
For this reason our study is not expected to give accurate information on 
the energy loss in the glasma. Nevertheless it seems relatively
safe to use the results for qualitative estimate of the glasma effect 
on the radiative energy loss.

Our results show that for RHIC and LHC conditions despite very large 
values of the coherent glasma color fields the order of magnitude 
of the synchrotron-like
energy loss turns out to be small as compared to 
the radiative energy loss in the QGP phase. 
For this reason the glasma phase 
should not play a significant role in the suppression of high-$p_{T}$
hadrons.
But due to a wide angle distribution the synchrotron-like gluon 
emission may potentially be important for the jet structure
at large angles (such as ridge effect and conical emission).
However, an accurate study of this region would require the calculations
beyond the small angle approximation.

\vspace {.7 cm}
\noindent
{\large\bf Acknowledgements}

\noindent
We are grateful to T.~Lappi for a useful communication on his
numerical simulation of the glasma.
One of the authors (BGZ) would like to thank LAPTH and CERN TH Division
for hospitality and partial support during his visits there.
The work of BGZ  is supported 
in part by the 
Laboratoire International Associ\'e "Physique Th\'eorique et Mati\`ere
Condens\'ee" (ENS-Landau), the grant RFBR
12-02-00063-a and the program
SS-6501.2010.2.

\newpage
%
%
\begin{figure} [h]
\begin{center}
\hspace{-1cm}
\epsfig{file=Fig1.eps,height=14cm,angle=0}
\end{center}
\caption[.]
{The ratio 
$R(q)=dN_{s}/d\omega d\qb\Big/
dN_{s}/d\omega d\qb\Big|_{\qb=0}$ 
for $q\to gq$ process
at $E_{q}=50$ GeV obtained using (\ref{eq:160})
for RHIC (left) and LHC (right) conditions for the gluon
energies $\omega=$2 (solid lines), 5 (short dashed lines), 10 
(long dashed lines), and 25 (dotted lines) GeV.
The transverse momentum vector is given by $\qb=(q,0)$ (upper panels),
and $\qb=(0,q)$ (lower panels), here $x$-axis in $\qb$-plane is parallel
to the Lorentz force $\fb$, and $y$-axis is perpendicular to it.
}
\end{figure}

\begin{figure}[t]
\begin{center}
\epsfig{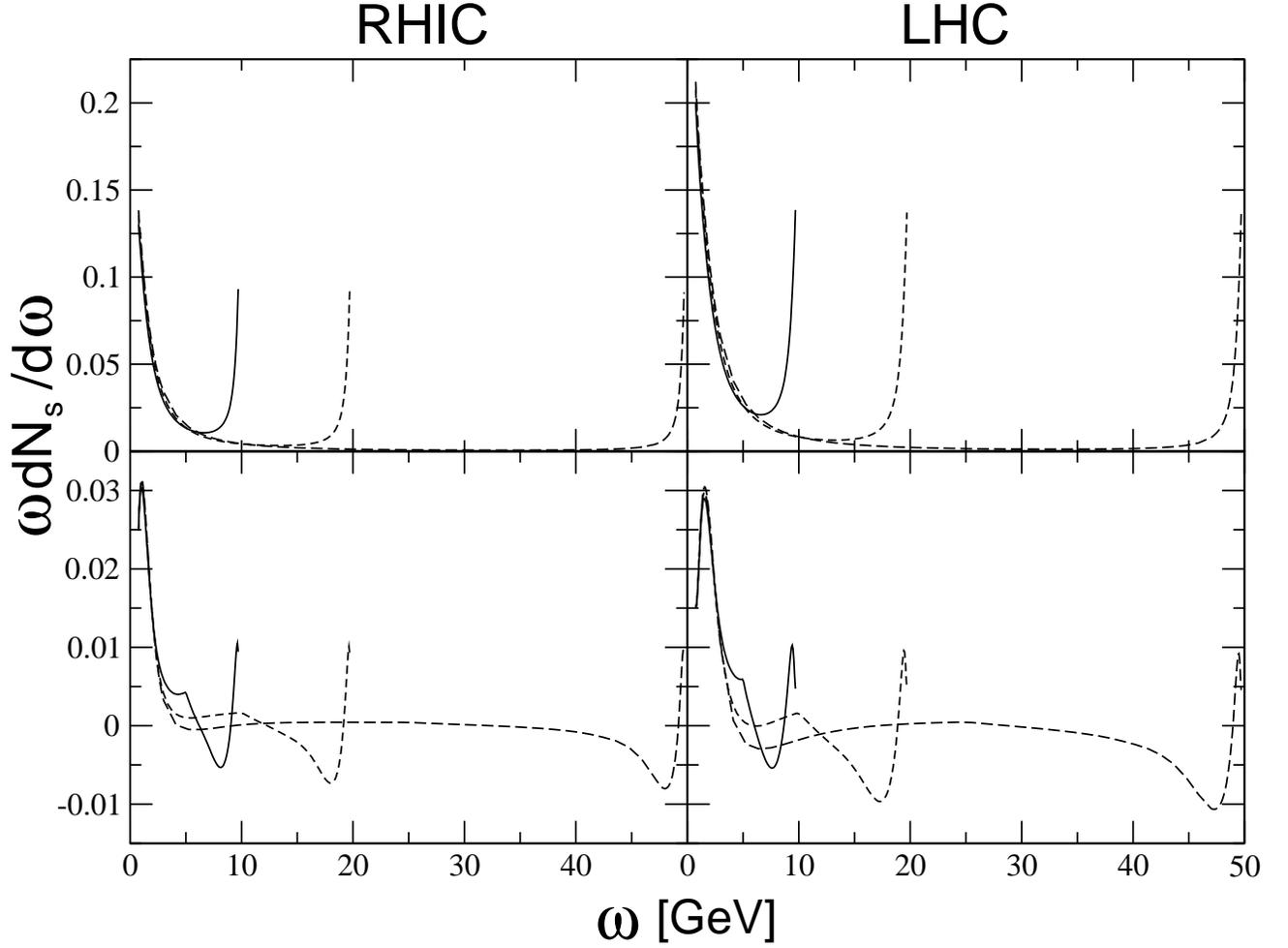}
\end{center}
\caption[.]
{
The energy loss spectrum $\omega dN_{s}/d\omega$ for $q\to gq$
process at $E_{q}=10$ (solid lines), 20 (short dashed lines), and 50 
(long dashed lines) GeV
for RHIC (left) and LHC (right) conditions
obtained from (\ref{eq:160}) without kinematical constraint on the transverse
momentum (upper panels) and with the restriction 
$|\qb|<\mbox{min}(\omega, E-\omega)$ (lower panels).
}
\end{figure}

\begin{figure}[t]
\begin{center}
\epsfig{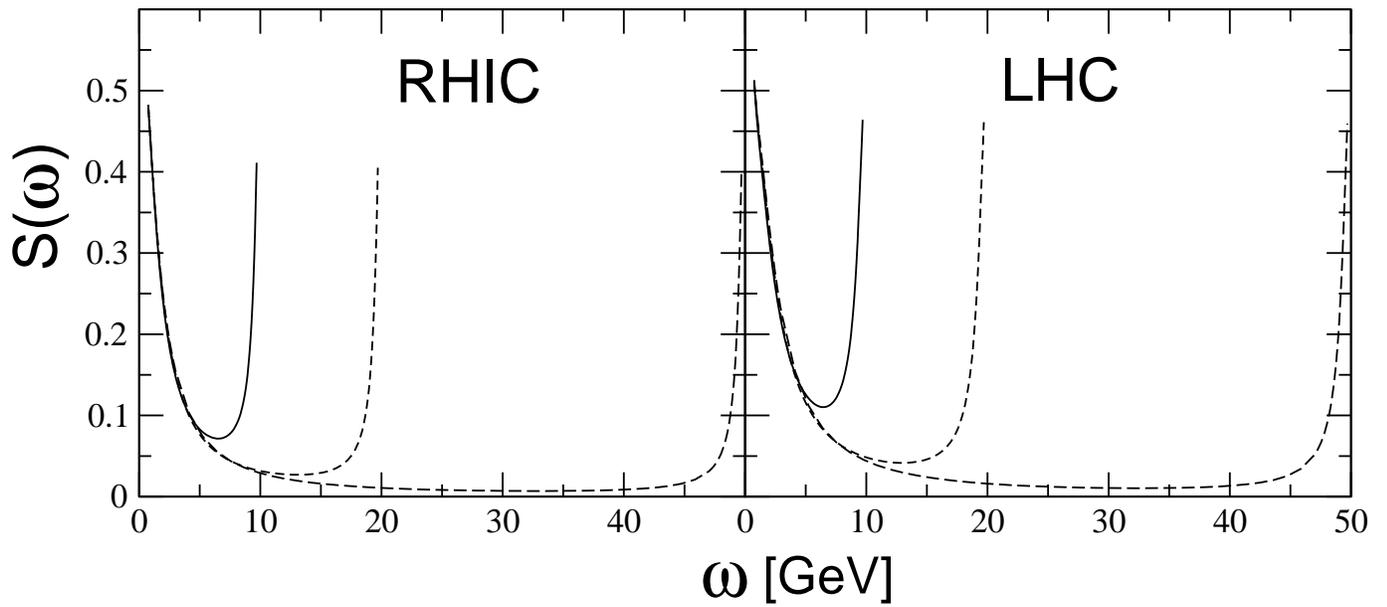}
\end{center}
\caption[.]{The finite-size suppression factor
for $q\to gq$ process for RHIC (left) and LHC (right) 
conditions
for the $\omega$-distribution obtained 
without kinematical constraint on transverse momentum, see text
for details.
The curves are for $E_{q}=10$ (solid lines), 20 (short dashed lines), and 
50 (long dashed lines) GeV.
}
\end{figure}

\end{document}